\documentclass{aa}  
\usepackage{hyperref}
\usepackage{float}
\usepackage{tabularx}
\usepackage{multirow}
\usepackage{footnote}
\usepackage{subfigure}
\usepackage{ae,aecompl}
\usepackage{graphicx}   
\usepackage{amsmath}    
\usepackage{amssymb}    
\usepackage[switch]{lineno}

\begin{document}

   \title{The peculiar abundances of HE~1005--1439\thanks{Based [in part] on data collected at the Subaru Telescope, which is operated by the National Astronomical Observatory of Japan (NAOJ).}\thanks{Part of the data are retrieved from the JVO portal (\url{http://jvo.nao.ac.jp/portal}) operated by the NAOJ.}}

   \subtitle{A carbon-enhanced  extremely metal-poor star contaminated with products of both {\it{s}}-~and {\it{i}}-process nucleosynthesis}

   \author{Partha Pratim Goswami
          \inst{1,2}
          \and
          Aruna Goswami\inst{1}
          }

   \institute{Indian Institute of Astrophysics, Koramangala, Bangalore
    560034, India  \\
              \email{partha.pg@iiap.res.in; aruna@iiap.res.in}
         \and
         Pondicherry University, R.V. Nagar, Kalapet, 605014, Puducherry, India\\
             }

   \date{Received 12 July 2021; Accepted 08 September 2021}

  \abstract
   {Understanding the surface chemical composition of carbon-enhanced metal-poor (CEMP) stars with enhanced abundances of heavy elements remains problematic.}
   {One of the primary objectives is to investigate the origin of the peculiar abundance pattern observed in the carbon-enhanced extremely metal-poor (EMP) object HE~1005--1439, which is enriched with both {\it{s}}-process and {\it{i}}-process nucleosynthesis products and thus forms a new class of object with a distinct abundance pattern.}
   {We performed a detailed, high-resolution spectroscopic analysis 
of this object based on SUBARU/HDS spectra with a resolution  R of $\sim$ 50~000.
We utilised the line analysis method with measured equivalent widths of neutral and ionised lines due to various elements. Moreover, we calculated the spectrum synthesis of carbon molecular bands and lines due to elements with hyperfine structures to determine the elemental abundances. Abundances of ten light elements from C through Ni and 12 heavy elements Sr, Y, Ba, La, Ce, Pr, Nd, Eu, Dy, Er, Hf,
 and Pb were determined. We also performed a parametric-model-based analysis of the abundances of the heavy elements to understand the origin of the observed abundance pattern.}
   {For the first time, we came across an object with a surface chemical composition that exhibits  contributions from both slow ({\it{s}}) and intermediate ({\it{i}}) neutron-capture nucleosynthesis. The  observed abundance pattern  is quite unique and has never been observed  in any CEMP stars. The star is found to be a CEMP-s star based on the CEMP stars' classification criteria. However, the observed abundance pattern could not be explained based on theoretical  {\it{s}}-process model predictions. On the contrary, our parametric-model based analysis clearly indicates its surface chemical composition being influenced by  similar contributions from both the {\it{s}}- and {\it{i}}-process.  We critically examined the observed abundances and carefully investigated the formation scenarios involving {\it{s}}-process and {\it{i}}-process that are available in literature, and we found that none of them could explain the observed abundances. We note that the variation we see  in our radial velocity estimates obtained from several epochs may  indicate the presence of a binary companion. Considering a binary system, we  therefore propose a formation scenario for this object involving effective proton ingestion episodes (PIEs) triggering {\it{i}}-process nucleosynthesis followed by {\it{s}}-process asymptotic giant branch (AGB) nucleosynthesis with a few third-dredge-up (TDU) episodes in the now extinct companion AGB star. Results obtained from the parametric-model-based analysis are  discussed in light of existing stellar evolutionary models.}
   {}


   \maketitle
%

\section{Introduction}
\label{sec:introduction}
Carbon-enhanced metal-poor (CEMP) stars form an important class of metal-poor giants, sub-giants, and dwarfs, with a large fraction of them showing enhanced abundances of heavy elements (see \citet{beers2005discovery} and  \citet{Frebel_review_2018} for a general review). Among the different types of CEMP stars, the CEMP-s stars are enriched with products of {\it{s}}-process nucleosynthesis, the CEMP-r stars are enriched with the products of {\it{r}}-process nucleosynthesis, and CEMP-r/s stars are enriched with products of {\it{i}}-process nucleosynthesis. Understanding the diverse abundance patterns exhibited by different groups of CEMP stars that are believed to be associated with different formation mechanisms has been a challenge. In \citet{Goswami_et_al_1_2021}, we present a detailed analysis and discussion on the classification criteria of CEMP stars, as well as the formation scenarios of CEMP stars put forward by different authors \citep{cowan1977,  hill2000heavy, qian2003stellar, cohen2003abundance, jonsell2006, Campbell_&_Lattanzio_2008, Campbell_et_al_2010, Stancliffe2011, Herwig_et_al_2011, Doherty_et_al_2015, abate2016cemp-rs, Jones_et_al_2016, Bannerjee_et_al_2018, Clarkson_et_al_2018, Denissenkov_et_al_2017, Cote_et_al_2018}.  In this paper, we report an extremely metal-poor carbon-enhanced star, HE~1005--1439, whose  surface chemical composition is found to be enriched with both {\it{s}}-process and {\it{i}}-process nucleosynthesis that forms a new class of object with a distinct abundance pattern. The peculiar abundance pattern, observed for the first time in a CEMP star, was investigated based on a parametric-model-based analysis that revealed almost equal contributions from both the  {\it{s}}-process and the {\it{i}}-process to its surface chemical composition. We examined various production mechanisms and formation scenarios for this object. A formation scenario involving effective proton ingestion episodes (PIEs) triggering {\it{i}}-process nucleosynthesis followed by {\it{s}}-process asymptotic giant branch (AGB) nucleosynthesis with limited third-dredge-up (TDU) episodes seems to be most promising for this type of object.
 
Literature surveys show that this object had been studied earlier by different groups \citep{aoki2007carbon, Schuler_et_al_2008, Yong_et_al_2013, Caffau_et_al.2017}. However, these studies were limited by the number of elements for which abundances were estimated. As the abundances of the neutron-capture elements except Ba are not reported in the literature, we re-visited the object and estimated the atmospheric parameters as well as abundances of ten light elements and twelve heavy elements.
The derived abundances were then carefully investigated with an aim to understand the origin and formation mechanism(s) of the object.

\section{Source of spectra}
\label{sec:source_of_spectra}
\begin{figure}
        \centering
        \includegraphics[height=7cm,width=9cm]{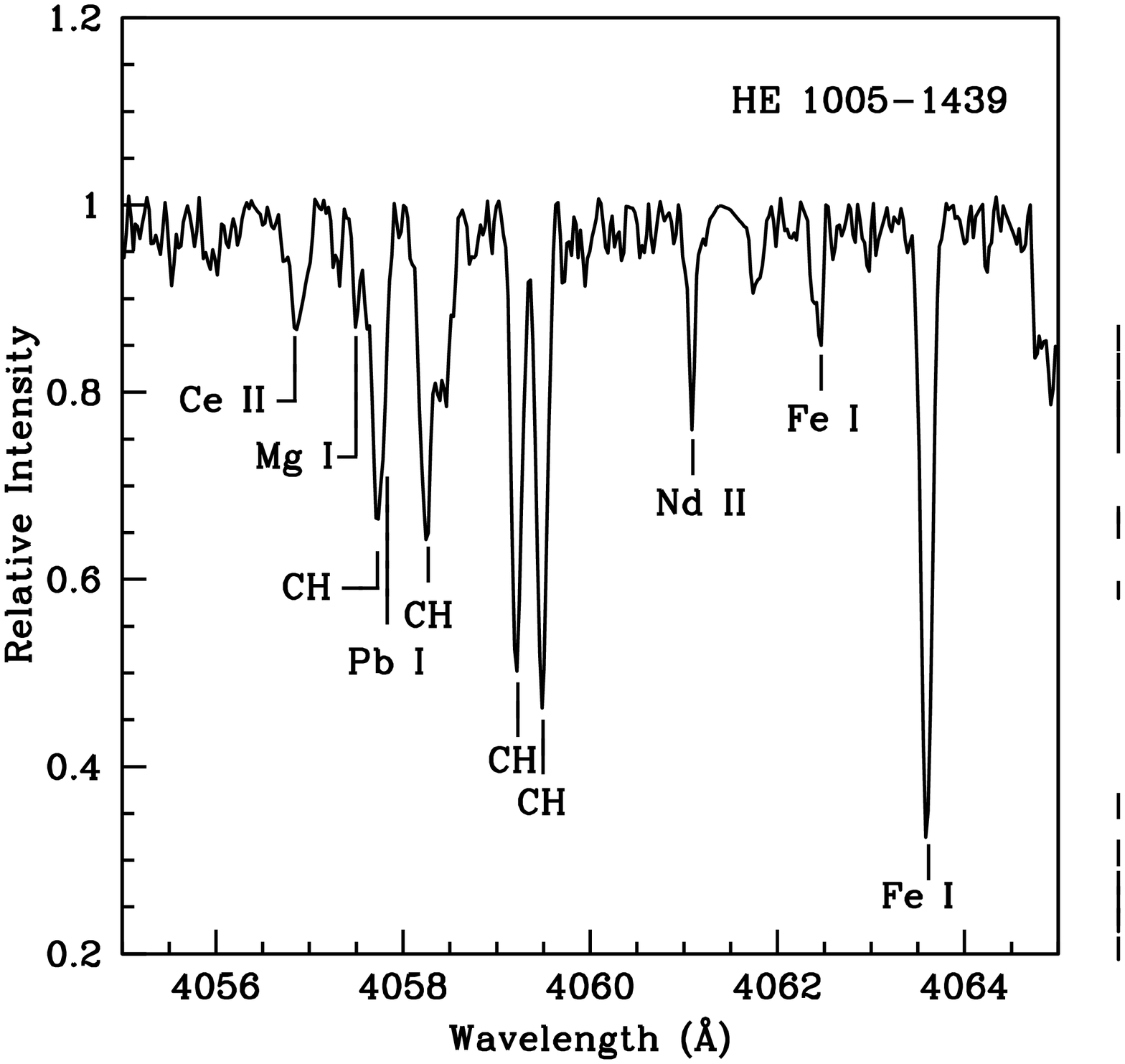}
                \includegraphics[height=7cm,width=9cm]{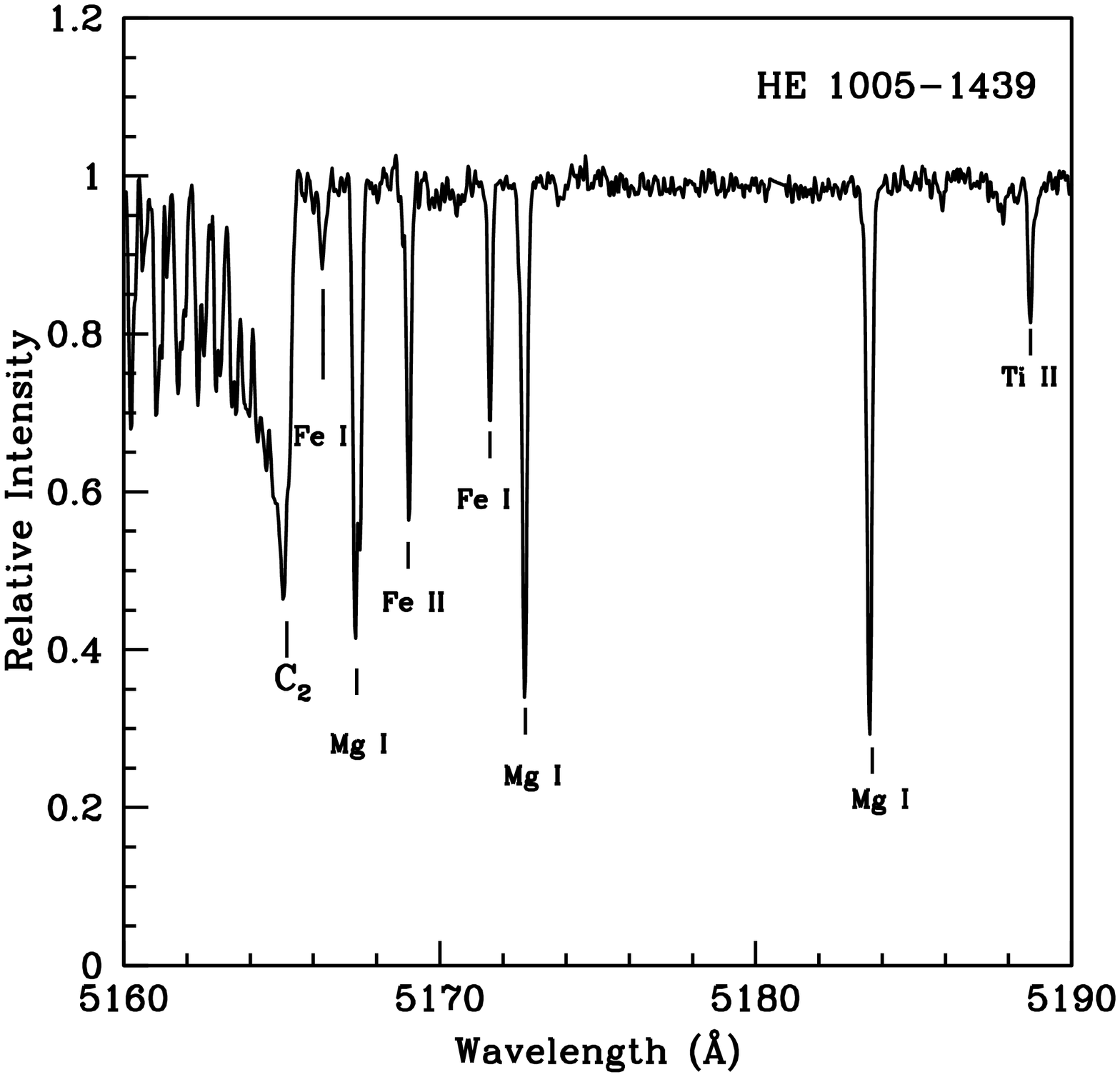}
        \caption{Sample spectra of the programme star in the 4055--4065 {\rm \AA} (upper panel) and 5160--5190 {\rm \AA} (bottom panel) wavelength regions.}
\label{fig:sample_spectra}
\end{figure}

The wavelength-calibrated high-resolution (R $\sim$ 50~000) spectra of HE~1005--1439 used in this study are retrieved from the SUBARU archive \footnote{\url{http://jvo.nao.ac.jp/portal}}. Spectra  obtained at four different epochs using the high-dispersion spectrograph (HDS) \citep{Noguchi_et_al_2002} attached to the 8.2m Subaru Telescope cover the 3515--6780 {\rm \AA\ wavelength range}. The spectra  acquired on October 26, 2002 (single exposure), and December 8, 2003 (five exposures) cover 4020--6780 {\rm \AA} with a gap  of $\sim$ 70 {\rm \AA} from 5370--5440 {\rm \AA,} and the spectra acquired on October 28, 2002 (single exposure) and May 26, 2003 (three exposures) cover the 3515--5270 {\rm \AA} wavelength region with a gap of 15 {\rm \AA} from 4380--4395 {\rm \AA}. For our studies, spectra obtained on the same dates are combined to increase the signal-to-noise ratio (S/N). The sample spectra of the programme star at two different wavelength regions are shown in Figure~\ref{fig:sample_spectra}.

\section{Radial velocities and stellar atmospheric parameters}
We measured the radial velocity of HE~1005--1439 from spectra acquired at four different epochs using several clean unblended lines. The estimated radial velocities with  98.54 $\pm$ 0.75 km~s$^{-1}$  (October 26, 2002), 99.17 $\pm$ 0.90 km~s$^{-1}$ (October 28, 2002), 48.95 $\pm$ 0.46 km~s$^{-1}$ (May 26, 2003), and 103.53 $\pm$ 0.46 km~s$^{-1}$ (December 8, 2003)  show that the object is a radial velocity  variable, and that it is likely to be in a binary system.

We determined the photometric temperatures of HE~1005--1439 using broad-band colours with colour-temperature calibrations available for giants \citep{alonso1999effective} based on the infrared flux method (IRFM) following the procedure described in \citet{goswami2006,goswami2016}. The 2MASS photometric magnitudes for J, H, and K are taken from \citet{2MASS_2003}. The photometric temperatures corresponding to (B--V), (V--K), (J--H), and (J--K) at [Fe/H] $\sim$ --3.0 are found to be T$_{eff}$(B--V) = 4650 K, T$_{eff}$(V--K) = 5323 K, T$_{eff}$(J--H) = 4817 K, and T$_{eff}$(J--K) = 5017 K, respectively. As the photometric temperature T$_{eff}$(J--K) is independent of metallicity \citep{alonso1996, alonso1999effective}, we used T$_{eff}$(J--K) as an initial guess for estimating the spectroscopic temperature of HE~1005--1439 through an iterative process of selecting the appropriate model atmosphere.
 
Following  the detailed procedure described in \citet{Goswami_et_al_1_2021}, the stellar atmospheric parameters T$_{eff}$, log~$g$, micro-turbulent velocity ($\zeta$), and metallicity [Fe/H] of the programme star are derived using 45 clean lines of Fe~I and five lines of Fe~II. The excitation potentials of these lines are in the 0.0--5.0 eV range. The list of lines  with the measured equivalent widths and atomic line information is presented in  Table~\ref{tab:Elem_linelist1}. 

An updated version of MOOG software by Sneden  \citep{Sneden_1973_MOOG}  that assumes local thermodynamic equilibrium (LTE) conditions was used for our analysis. The model atmospheres are used from the Kurucz grid of model atmospheres with no convective overshooting \footnote{\url{http://kurucz.harvard.edu/grids.html}}. The solar abundances are adopted from \citep{asplund2009}. The effective temperature and the micro-turbulent velocity are fixed by the conventional methods of excitation potential balance and equivalent width balance. The ionisation equilibrium method of equating the abundance of Fe derived from neutral and ionised Fe lines determines the surface gravity. The metallicity is given by the abundance of Fe derived from Fe~I and Fe~II lines. The estimates of stellar parameters along with the literature values are listed in Table~\ref{tab:atm_paracomp}.

Our estimate of effective temperature is 170 K higher than that of \citet{aoki2007carbon}. \citet{aoki2007carbon} estimated the effective temperature from (V--K), (V--R), (V--I), and (R--I) photometric colour indices, using the empirical temperature calibration scale of \citet{alonso1999effective}. In general, due to the presence of C$_{2}$ and CN molecular bands in carbon enhanced stars, (V--K) gives lower temperature than the other colour indices. However, in the case of HE~1005--1439 temperature estimate from T$_{eff}$-(V--K) calibration relation is found to be much higher than those obtained from other colour indices. This was also noticed by \citet{aoki2007carbon} and due to this discrepancy, adopted a lower value for temperature ($\sim$ 5000 K) closer to those derived using different colour indices. \citet{Yong_et_al_2013} estimated the effective temperature of HE~1005--1439 using IRFM adopting the colour-temperature relations given in \citet{Ramirez_&_Melendez_2005}. Our estimate of effective temperature is close to that of \citet{Yong_et_al_2013}.

Our estimate of micro-turbulent velocity ($\zeta$) is lower than that of \citet{aoki2007carbon} and \citet{Yong_et_al_2013}. For their analysis, \citet{Yong_et_al_2013} used the same measured equivalent widths and lines reported by \citet{aoki2007carbon}. While \citet{aoki2007carbon} used 19 lines of Fe~I covering a range of equivalent widths 16.7 {\rm m\AA} -- 88.0 {\rm \AA}, we have used 45 lines of Fe~I covering a range of equivalent widths 15.6 {\rm \AA} -- 127.1 {\rm \AA}. With more clean lines covering a good range
in line strength, we are confident about our estimated micro-turbulent velocity.

Our measured surface gravity log~$g$ is similar to that of \citet{aoki2007carbon}. The log~$g$ derived by \citet{aoki2007carbon} is based on the ionisation equilibrium method similar to the one that we have used. \citet{Yong_et_al_2013} used Y$^{2}$ isochrones \citep{Demarque_et_al_2004} to determine log~$g$, assuming an age of 10~Gyr, and [$\alpha$/Fe] = +0.3 and reported a higher log~$g$ ($\sim$ 2.55) for this object.

We estimated the mass of HE~1005--1439 from its position in the Hertzsprung-Russell (HR) diagram (log(L/L$_{\odot}$) versus log(T$_{eff}$) plot). The value of the parallax (= 0.2733 mas) is taken from \citet{gaia2018} and the V (= 13.52) magnitude is taken from SIMBAD. The bolometric correction is determined based on the empirical calibration equations of \citet{alonso1999effective}. Interstellar extinction for HE~1005--1439 is calculated using the formulae by \citet{chen_et_al_1998}. The value of log(L/L$_{\odot}$) is found to be $\sim$ 1.83. We have used the updated BaSTI-IAC evolutionary tracks \footnote{\url{http://basti-iac.oa-abruzzo.inaf.it/}} \citep{Hidalgo_et_al_2018}  generated for [Fe/H] = --3.2 and [$\alpha$/Fe] = 0.4, including overshooting and diffusion, to estimate the mass of the star. The mass of the object is found to be 0.8 M$_{\odot}$. The log~$g$ value is calculated using the following relation:

\begin{equation}
log~(g/g_{\odot}) =  log~(M/M_{\odot}) + 4log~(T_{eff}/T_{eff\odot}) + 0.4(M_{bol} - M_{bol\odot})
.\end{equation}

The adopted solar values are log~$g\odot$ = 4.44, T$_{eff}\odot$ =5770 K and M$_{bol\odot}$ = 4.74 mag \citep{Yang_et_al_2016}. Using this method, we found a log~$g$ ($\sim$ 2.32) comparable to that of \citet{Yong_et_al_2013}. This value is much larger than our spectroscopic log~$g$ value $\sim$1.8. We note that the evolutionary tracks and isochrones highly depend on the opacity in the stellar atmosphere. As BaSTI-IAC evolutionary tracks are generated using normal carbon and without considering the influence of high carbon, log~$g$ values determined using such evolutionary tracks and isochrones may lead to erroneous estimates. This may also explain the discrepancy between the log~$g$ values obtained by us from spectroscopy and that reported by \citet{Yong_et_al_2013}. We used the spectroscopic log ~$g$ value for our  analysis.

{\footnotesize
\begin{table*}
\caption{\bf{Derived atmospheric parameters of our programme star, and literature values. }}
\label{tab:atm_paracomp} 
\begin{tabular}{lccccccccc}
\hline     
Star name       & T$_{eff}$  &log g  & $\zeta    $   & [Fe~I/H]           &  [Fe~II/H]         & [Fe/H]   & Ref \\
                &    (K)     & (cgs) & (km~s$^{-1}$) &                    &                    &          &     \\
\hline
HE~1005--1439   &   5170     &  1.80 &  1.26         & $-$3.04 $\pm$ 0.15 & $-$3.01 $\pm$ 0.03 & $-$3.03  & 1   \\
                &   5000     &  1.90 &  2.00         & $-$3.17 $\pm$ 0.32 & $-$3.15 $\pm$ 0.22 & $-$3.17  & 2   \\
                &   5202     &  2.55 &  1.60         &     -              &        -           & $-$3.09  & 3   \\
                &   5030     &  -    &   -           &     -              &        -           &   -      & 4   \\
\hline
\end{tabular}

References: 1. Our work; 2. \citet{aoki2007carbon}; 3. \citet{Yong_et_al_2013}; 4. \citet{gaia2018}.  \\          
\end{table*}
}

\section{Results \& discussions}

\subsection{Abundance analysis}
\label{sec:abundance_analysis}

We determined the elemental abundances by measuring the equivalent widths of the absorption lines due to neutral and ionised species of several elements and/or by applying a spectrum synthesis technique using the radiative transfer code MOOG \citep{Sneden_1973_MOOG} that assumes LTE and  model atmospheres  from the Kurucz grid of model atmospheres with no convective overshooting \footnote{\url{http://kurucz.harvard.edu/grids.html}}. Elemental abundances of C, Na, Mg, Ca, Sc, Ti, Cr, and Mn, iron-peak elements Co and Ni, and  neutron-capture elements  Sr, Y, Ba, La, Ce, Pr, Nd, Eu, Dy, Er, Hf, and Pb were estimated. We used the method of spectrum synthesis calculations for elements showing hyperfine splitting (e.g. Sc, Mn, Ba, La, and Eu). The lines used in the analysis with the measured equivalent widths and atomic line information are presented in Table~\ref{tab:Elem_linelist1}. Atomic line information, such as the excitation potential and log~$gf$ values, were taken from the Kurucz database of atomic line list. The abundance results along with the literature values are presented in Table~\ref{tab:abundances}.

\subsubsection{Light elements}

\begin{figure}
        \centering
        \includegraphics[height=7cm,width=9cm]{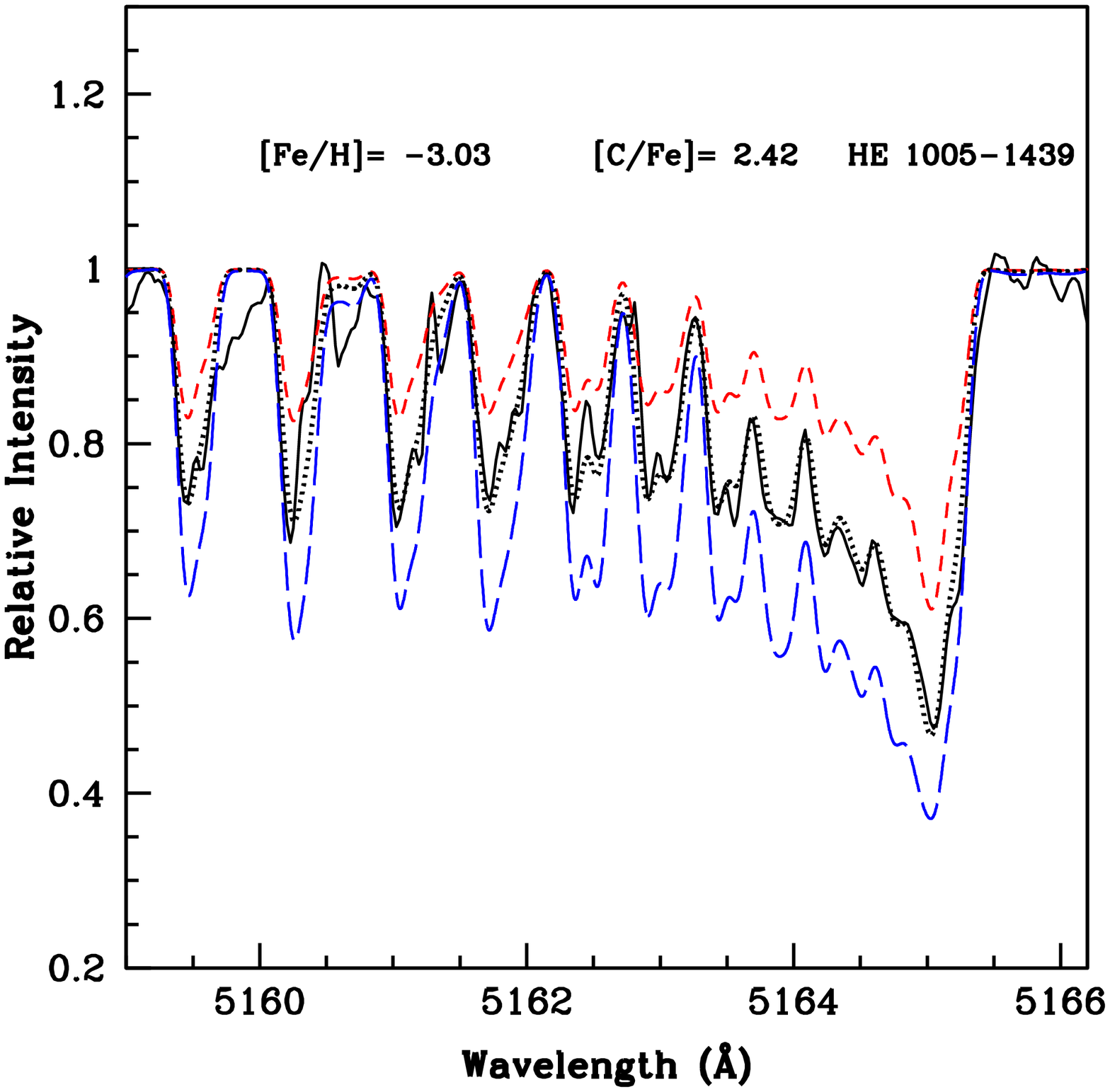}
                \includegraphics[height=7cm,width=9cm]{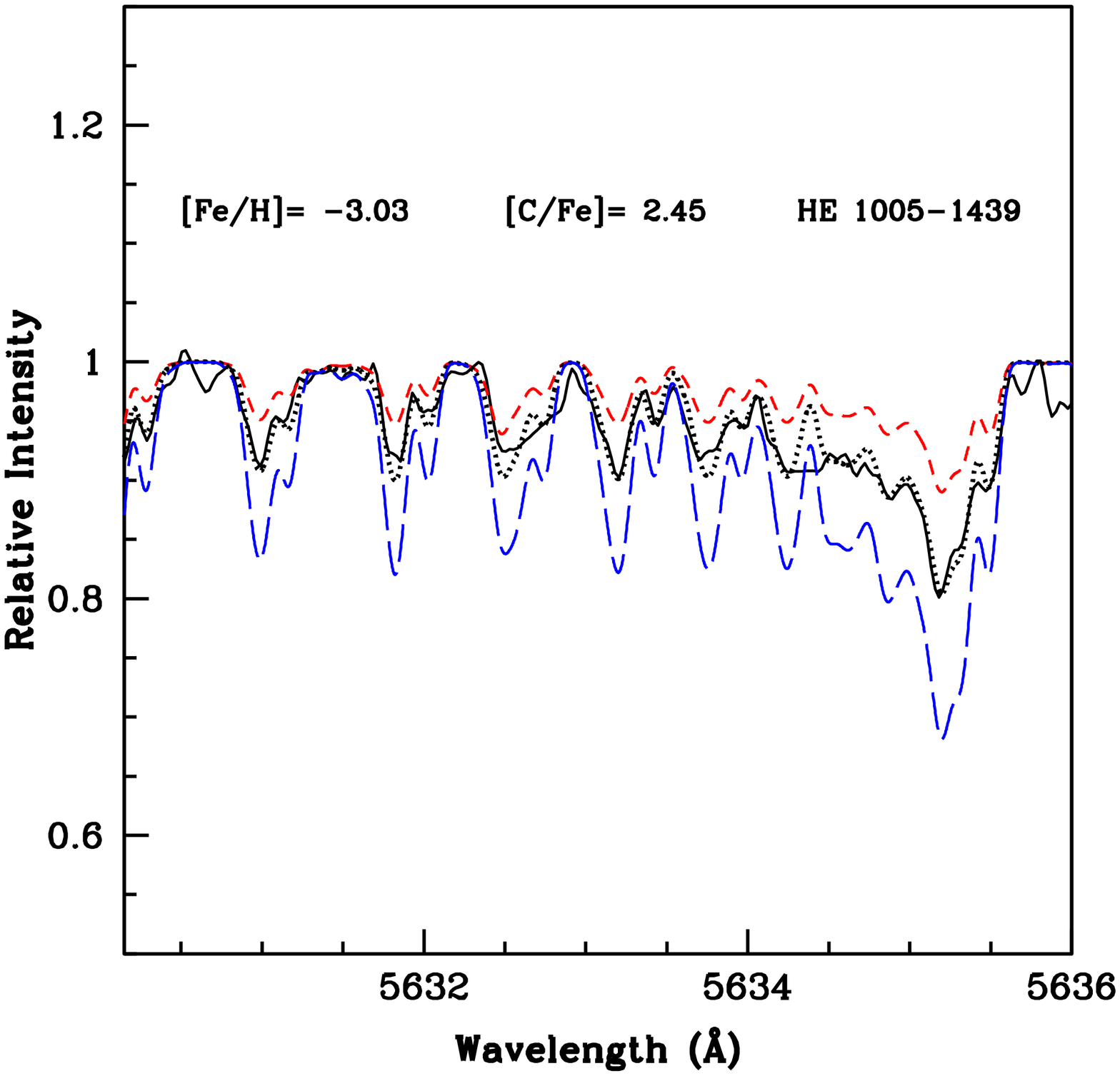}
        \caption{Spectral synthesis fits of C$_{2}$ bands around 5165 {\rm \AA} (top panel) and 5635 {\rm \AA} (bottom panel). The dotted lines indicate the synthesised spectra and the solid lines indicate the observed spectra. Two alternative synthetic spectra are shown corresponding to $\Delta$[C/Fe] = +0.3 (long-dashed line) and $\Delta$[C/Fe] = $-$0.3 (short-dashed line).}
\label{fig:C2}
\end{figure}

\begin{figure}
        \centering
        \includegraphics[height=7.5cm,width=9cm]{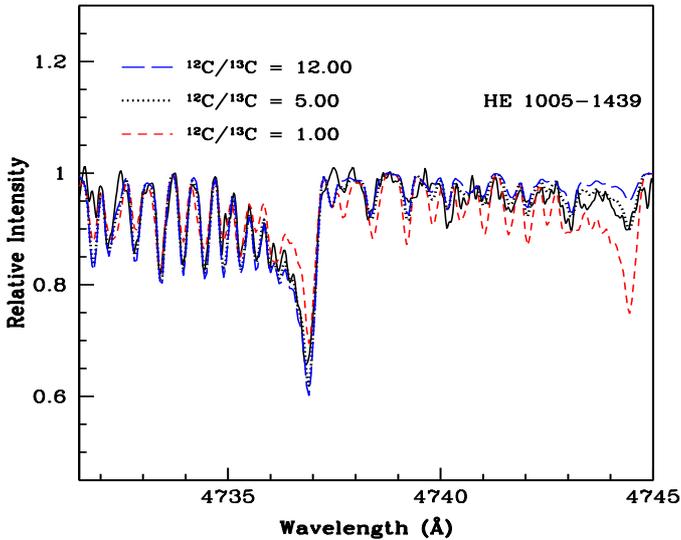}
        \caption{Spectral synthesis fits (dotted curves) of the C$_{2}$ features around 4740 {\rm \AA} obtained with the adopted C abundance and $^{12}$C/$^{13}$C value (dotted curve). The observed spectrum is shown by a solid curve. Two alternative fits with $^{12}$C/$^{13}$C $\sim$ 1 (short-dashed line) and 12 (long-dashed line) are shown to illustrate the sensitivity of the line strengths to the isotopic carbon abundance ratios.}
\label{fig:C12C13}
 \end{figure}

We used spectrum synthesis calculations for estimating the abundances of carbon from molecular bands of carbon. 
The abundance of carbon was derived using the CH band near 4310 {\rm \AA} and C$_{2}$ bands near 5160 {\rm \AA} and 5635 {\rm \AA} (Figure~\ref{fig:C2}). The slight difference in A(C) derived from C$_{2}$ and CH molecular bands might have appeared due to the difference in S/N in the regions of the CH and C$_{2}$ bands. The S/N of the spectra near 5200 {\rm \AA} (i.e. near the C$_{2}$ bands) is $\sim$ 130, while the S/N is $\sim$ 50 near 4320 {\rm \AA} (i.e. near the CH band). Carbon is found to be enhanced in HE~1005--1439 with [C/Fe]~$\sim$~2.37. 
The estimated carbon isotopic ratio  $^{12}$C/$^{13}$C, obtained from  spectrum synthesis calculations of the C$_{2}$ band near 4740 {\rm \AA} (Figure~\ref{fig:C12C13}) is ${\sim}$~5.0. We have taken the line lists of C$_{2}$ bands at 5165 {\rm \AA}, 5635 {\rm \AA,} 
and 4740 {\rm \AA,} and of the CH band at 4310 {\rm \AA} from the `linemake'\footnote{{\textit{linemake}} contains laboratory atomic data (transition probabilities, hyperfine and isotopic substructures) published by the Wisconsin Atomic Physics and the Old Dominion Molecular Physics
groups. These lists and accompanying line list assembly software
have been developed by C. Sneden and are curated by V. Placco
at https://github.com/vmplacco/linemake.} atomic and molecular line database.

While HE~1005--1439 is enhanced in Na ([Na/Fe]~=~1.37), Mg ([Mg/Fe]~=~0.53) is moderately enhanced. \citet{Gehren_et_al_2004} found that  the systematic NLTE corrections required on the abundances of Na and Mg in case of metal-poor halo stars are --0.4 and +0.1, respectively.

The other light elements, Ca, Sc, Ti, Co, and Ni, are moderately enhanced in HE~1005--1439. The abundances of Cr and Mn are found to be sub-solar. However, \citet{Bergemann_&_Gehren_2008} and \citet{Bergemann_et_al_2010} have found that in case of metal-poor giants, abundances of Co and Mn are underestimated in LTE calculations. \citet{Bergemann_&_Gehren_2008} performed NLTE calculations for Mn on a sample of fourteen stars and found that the NLTE abundances of Mn in all the sample stars are higher than the LTE abundances, in fact, the NLTE correction may go up to 0.5 -- 0.7 dex at low metallicities. NLTE calculations done by \citet{Bergemann_et_al_2010} for Co I lines revealed that the corrections may vary from 0.1 -- 0.8 dex depending on the effective temperature and metallicity. They found that deviation from LTE is larger in the case of giants than it is for dwarfs. Our estimates of the  abundances of light elements are found to match closely  with the literature values (calculated with LTE assumption) of the programme star within error bars. Hyperfine splitting contributions of the lines (Table~\ref{tab:Elem_linelist1}) used for spectrum synthesis calculations of Sc and Mn are taken from linemake.

\subsubsection{Neutron-capture elements}

While HE~1005--1439 is found to be enhanced ([X/Fe]~$>$~1.0) in heavy elements Ba, La, Ce, Pr, Nd, Er, Hf, and Pb, the elements Sr, Y, Eu, and Dy are moderately enhanced ([X/Fe] in the range 0.26 to 0.72). We could not detect lines due to Tc. Hyperfine splitting contributions for the spectrum synthesis calculations of the lines Ba~II~5853.67~{\rm \AA} \& Ba~II~6141.71~{\rm \AA} are taken from \citet{mcwilliam1998} and for La~II~4921.78~{\rm \AA} line the hyperfine splitting contributions are taken from \citet{jonsell2006}. For the spectrum synthesis calculations of the lines Ba~II~6496.90~{\rm \AA}, La~II~4086.71~{\rm \AA}, La~II~4123.22~{\rm \AA}, Eu~II~4129.73~{\rm \AA,} and Eu~II~4205.04~{\rm \AA,} the hyperfine splitting contributions are taken from linemake. The abundance of Pb is estimated using the spectrum synthesis calculation of Pb~I~4057.81~{\rm \AA}.

\subsection{Abundance uncertainties}

We estimated the total uncertainties on the elemental abundances as discussed in \citet{Goswami_et_al_1_2021}. Two components, namely random error ($\sigma_{ran}$ = $\dfrac{\sigma_{s}}{\sqrt N}$, where $\sigma_{s}$ represents the standard deviation of the abundance of a particular species derived using N number of lines of that species) and systematic error ($\sigma_{sys}$), contribute to the total uncertainties. While the random error arises due to the uncertainties on the factors like oscillator strength, equivalent width measurement, and line blending, the systematic error arises due to the uncertainties in estimating the stellar atmospheric parameters. Finally, the uncertainties on [X/Fe] are derived as follows:

\begin{equation}
\sigma ^{2}_{[X/Fe]} =  \sigma ^{2}_{log\epsilon} + \sigma ^{2}_{[Fe/H]}
,\end{equation}

\begin{equation}
\sigma ^{2}_{log\epsilon} =  \sigma ^{2}_{ran} + \sigma ^{2}_{sys}
,\end{equation}

\begin{equation}
\label{eqn:uncertainty}
\begin{split}
\sigma ^{2}_{sys} = & \left( \dfrac{\delta log\epsilon}{\delta T}\right)^{2}\sigma ^{2}_{T_{eff}} + \left( \dfrac{\delta log\epsilon}{\delta logg}\right)^{2}\sigma ^{2}_{logg}\\
                            & + \left( \dfrac{\delta log\epsilon}{\delta \zeta}\right)^{2}\sigma ^{2}_{\zeta} + \left( \dfrac{\delta log\epsilon}{\delta [Fe/H]}\right)^{2}\sigma ^{2}_{[Fe/H]} 
\end{split}
,\end{equation}

where, $\sigma_{T_{eff}}$ = 100 K, $\sigma_{logg}$ = 0.2 dex, $\sigma_{\zeta}$ = 0.2 km/s$^{-1}$, and $\sigma_{[Fe/H]}$ = 0.15 dex represent the typical uncertainties on the stellar atmospheric parameters T$_{eff}$, logg, $\zeta$, and [Fe/H], respectively. We evaluated the partial derivatives appearing in Equation~\ref{eqn:uncertainty} for the programme star, varying the stellar parameters T$_{eff}$, logg, $\zeta,$ and [Fe/H] by $\pm$ 100 K, $\pm$ 0.2 dex, $\pm$ 0.2 km/s$^{-1}$, and $\pm$ 0.2 dex, respectively. We note that the uncorrelated nature of the uncertainties arising from the different stellar parameters in Equation~\ref{eqn:uncertainty} may lead to the overestimation of the calculated uncertainties on log~$\epsilon$ and [X/Fe]. The estimated uncertainties $\sigma_{log{\epsilon}}$ and $\sigma_{[X/Fe]}$ are listed in columns 5 and 8, respectively, of Table~\ref{tab:abundances}.

{\footnotesize
\begin{table*}
\caption{\bf{Elemental abundances in HE~1005--1439}}
\label{tab:abundances}
\scalebox{0.90}{
\begin{tabular}{l c c|c c c c c|c|c|c}
\hline
\hline
            & Z  & solar $log{\epsilon}^{(a)}$ & $log{\epsilon}$&$\sigma_{log{\epsilon}}$&[X/H]&[X/Fe]&$\sigma_{[X/Fe]}$&[X/Fe]&[X/Fe]&[X/Fe]      \\
            &    &                             &    (dex)       &                       &     &      &              & $^{(b)}$&$^{(c)}$&$^{(d)}$\\
\hline
C (CH, 4310 {\rm \AA})      & 6  & 8.43 & 7.65 (syn)          & 0.23  & $-$0.78    & 2.25    & 0.27 &  2.48    &   -       &  2.14$^{**}$ \\
C (C$_{2}$, 5165 {\rm \AA}) & 6  & 8.43 & 7.82 (syn)          & 0.22  & $-$0.61    & 2.42    & 0.27 &   -      &   -       &   -         \\
C (C$_{2}$, 5635 {\rm \AA}) & 6  & 8.43 & 7.85 (syn)          & 0.21  & $-$0.58    & 2.45    & 0.26 &   -      &   -       &   -         \\
Na {\sc i}                  & 11 & 6.24 & 4.58$\pm$0.03 (2)   & 0.16  & $-$1.66    & 1.37    & 0.22 &  1.19    &  1.05     &   -         \\
Mg {\sc i}                  & 12 & 7.60 & 5.10$\pm$0.05 (4)   & 0.12  & $-$2.50    & 0.53    & 0.19 &  0.60    &  0.33     &   -         \\
Ca {\sc i}                  & 20 & 6.34 & 3.78$\pm$0.18 (10)  & 0.10  & $-$2.56    & 0.47    & 0.18 &  0.57    &  0.54     &   -         \\
Sc {\sc ii}                 & 21 & 3.15 & 0.82$\pm$0.10 (2)   & 0.17  & $-$2.33    & 0.70    & 0.23 &   -      &   -       &   -         \\ 
Sc {\sc ii}$^{*}$           & 21 & 3.15 & 0.75$\pm$0.00 (2)   & 0.15  & $-$2.40    & 0.63    & 0.23 &   -      &   -       &   -         \\ 
Ti {\sc i}                  & 22 & 4.95 & 2.27$\pm$0.17 (4)   & 0.14  & $-$2.68    & 0.35    & 0.21 &  0.48    &  0.49     &   -         \\
Ti {\sc ii}                 & 22 & 4.95 & 2.23$\pm$0.20 (8)   & 0.12  & $-$2.72    & 0.31    & 0.20 &  0.19    &  0.31     &   -         \\
Cr {\sc i}                  & 24 & 5.64 & 2.59$\pm$0.15 (5)   & 0.17  & $-$3.05    & $-$0.02 & 0.23 & $-$0.38  & $-$0.34   &   -         \\
Mn {\sc i}                  & 25 & 5.43 & 2.56$\pm$0.14 (2)   & 0.25  & $-$2.87    & 0.16    & 0.29 &   -      &   -       &   -         \\
Mn {\sc i}$^{*}$            & 25 & 5.43 & 2.13$\pm$0.03 (2)   & 0.23  & $-$3.30    & $-$0.27 & 0.28 &   -      &   -       &   -         \\
Fe {\sc i}                  & 26 & 7.50 & 4.46$\pm$0.15 (45)  & 0.15  & $-$3.04    & -       &   -  &   -      &   -       &   -         \\
Fe {\sc ii}                 & 26 & 7.50 & 4.49$\pm$0.03 (5)   & 0.10  & $-$3.01    & -       &   -  &   -      &   -       &   -         \\
Co {\sc i}                  & 27 & 4.99 & 2.20$\pm$0.07 (3)   & 0.13  & $-$2.79    & 0.24    & 0.20 &   -      &   -       &   -         \\
Ni {\sc i}                  & 28 & 6.22 & 3.40 (1)            & 0.23  & $-$2.82    & 0.21    & 0.28 &   -      &   -       &   -         \\
Sr {\sc ii}                 & 38 & 2.87 & 0.10$\pm$0.22 (2)   & 0.27  & $-$2.77    & 0.26    & 0.31 &   -      &   -       &   -         \\
Y {\sc ii}                  & 39 & 2.21 &$-$0.40$\pm$0.27 (3) & 0.18  & $-$2.61    & 0.42    & 0.23 &   -      &   -       &   -         \\
Ba {\sc ii}                 & 56 & 2.18 & 0.64$\pm$0.27 (3)   & 0.28  & $-$1.54    & 1.49    & 0.31 &   -      &   -       &   -         \\
Ba {\sc ii}$^{*}$           & 56 & 2.18 & 0.31$\pm$0.24 (3)   & 0.27  & $-$1.87    & 1.16    & 0.31 & 1.06     &  1.17     &   -         \\
La {\sc ii}                 & 57 & 1.10 &$-$0.61$\pm$0.12 (3) & 0.13  & $-$1.71    & 1.32    & 0.20 &   -      &   -       &   -         \\
La {\sc ii}$^{*}$           & 57 & 1.10 &$-$0.68$\pm$0.06 (3) & 0.11  & $-$1.78    & 1.25    & 0.19 &   -      &   -       &   -         \\
Ce {\sc ii}                 & 58 & 1.58 &$-$0.07$\pm$0.17 (3) & 0.14  & $-$1.65    & 1.38    & 0.21 &   -      &   -       &   -         \\
Pr {\sc ii}                 & 59 & 0.72 &$-$0.76 (1)          & 0.22  & $-$1.48    & 1.55    & 0.27 &   -      &   -       &   -         \\
Nd {\sc ii}                 & 60 & 1.42 &$-$0.36$\pm$0.15 (4) & 0.13  & $-$1.78    & 1.25    & 0.20 &   -      &   -       &   -         \\
Eu {\sc ii}$^{*}$           & 63 & 0.52 &$-$2.05$\pm$0.18 (2) & 0.17  & $-$2.57    & 0.46    & 0.22 &   -      &   -       &   -         \\
Dy {\sc ii}                 & 66 & 1.10 &$-$1.21$\pm$0.19 (3) & 0.15  & $-$2.31    & 0.72    & 0.21 &   -      &   -       &   -         \\
Er {\sc ii}                 & 68 & 0.92 &$-$1.05$\pm$0.20 (2) & 0.17  & $-$1.97    & 1.06    & 0.23 &   -      &   -       &   -         \\
Hf {\sc ii}                 & 72 & 0.85 &$-$0.73$\pm$0.01 (2) & 0.16  & $-$1.58    & 1.45    & 0.18 &   -      &   -       &   -         \\
Pb {\sc ii}$^{*}$           & 82 & 1.75 &   0.70 (1)          & 0.11  & $-$1.05    & 1.98    & 0.19 &   -      &   -       &   -         \\
\hline 
\\
 $^{12}$C/$^{13}$C  (C$_{2}$, 4740 {\rm \AA}) = 5.0\\

\hline
 \end{tabular}}

$^{*}$ abundance is derived using spectrum synthesis calculations. 
$^{**}$ abundance is derived using [C {\sc i}] line. 
The number inside the parenthesis shows the number of lines used for the abundance determination.\\ \textbf{References:} $^{(a)}$ \citet{asplund2009}; $^{(b)}$\citet{aoki2007carbon}; $^{(c)}$\citet{Yong_et_al_2013}; $^{(d)}$\citet{Schuler_et_al_2008}.\\
\end{table*}
}

\subsection{Interpretation of results}
\begin{figure*}
     \begin{center}
\centering
        {%
                 \label{fig:laeubaeu}
            \includegraphics[height=6.5cm,width=7cm]{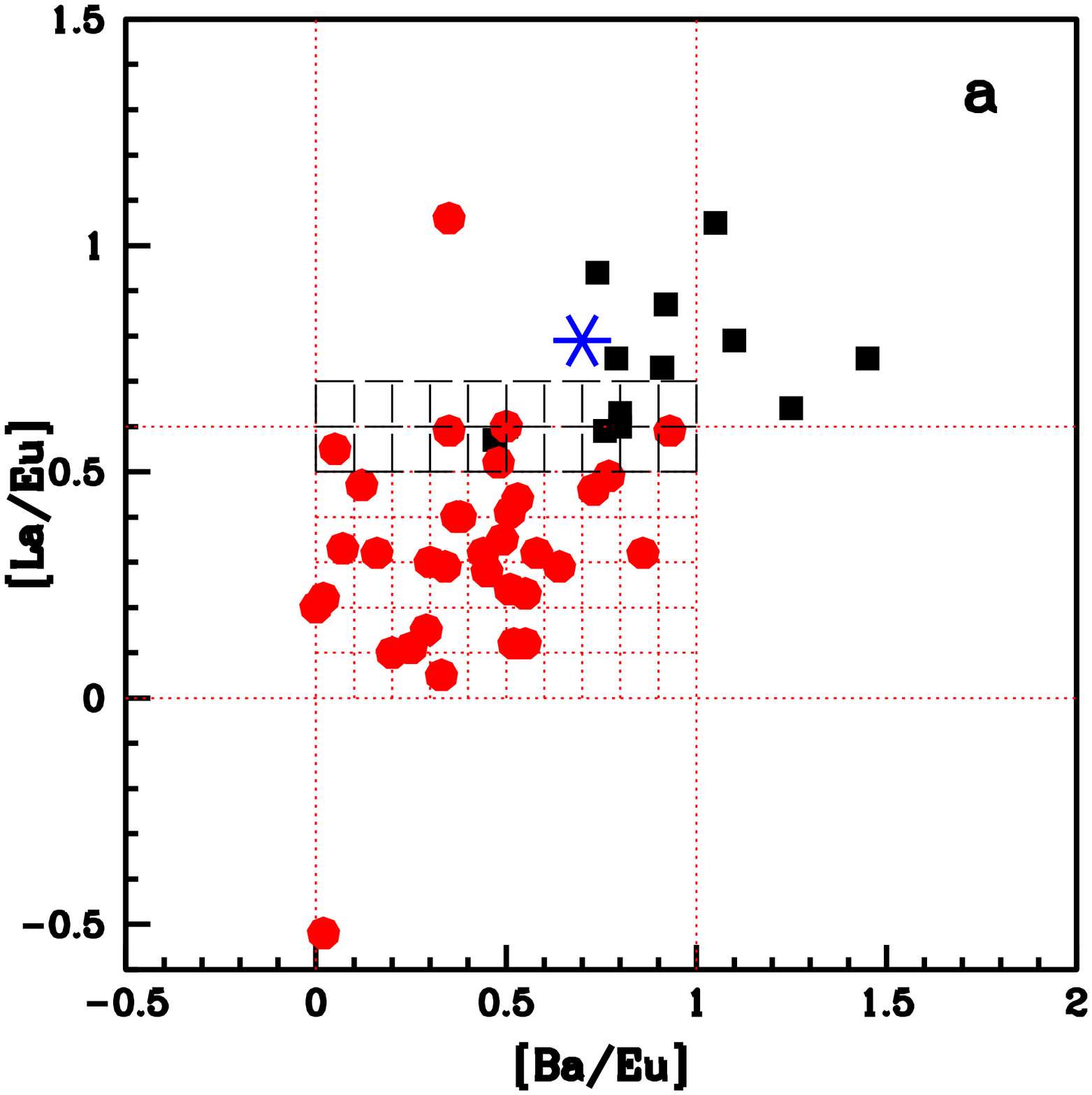}
        }%
        {%
      \label{fig:hs}
            \includegraphics[height=6.5cm,width=7cm]{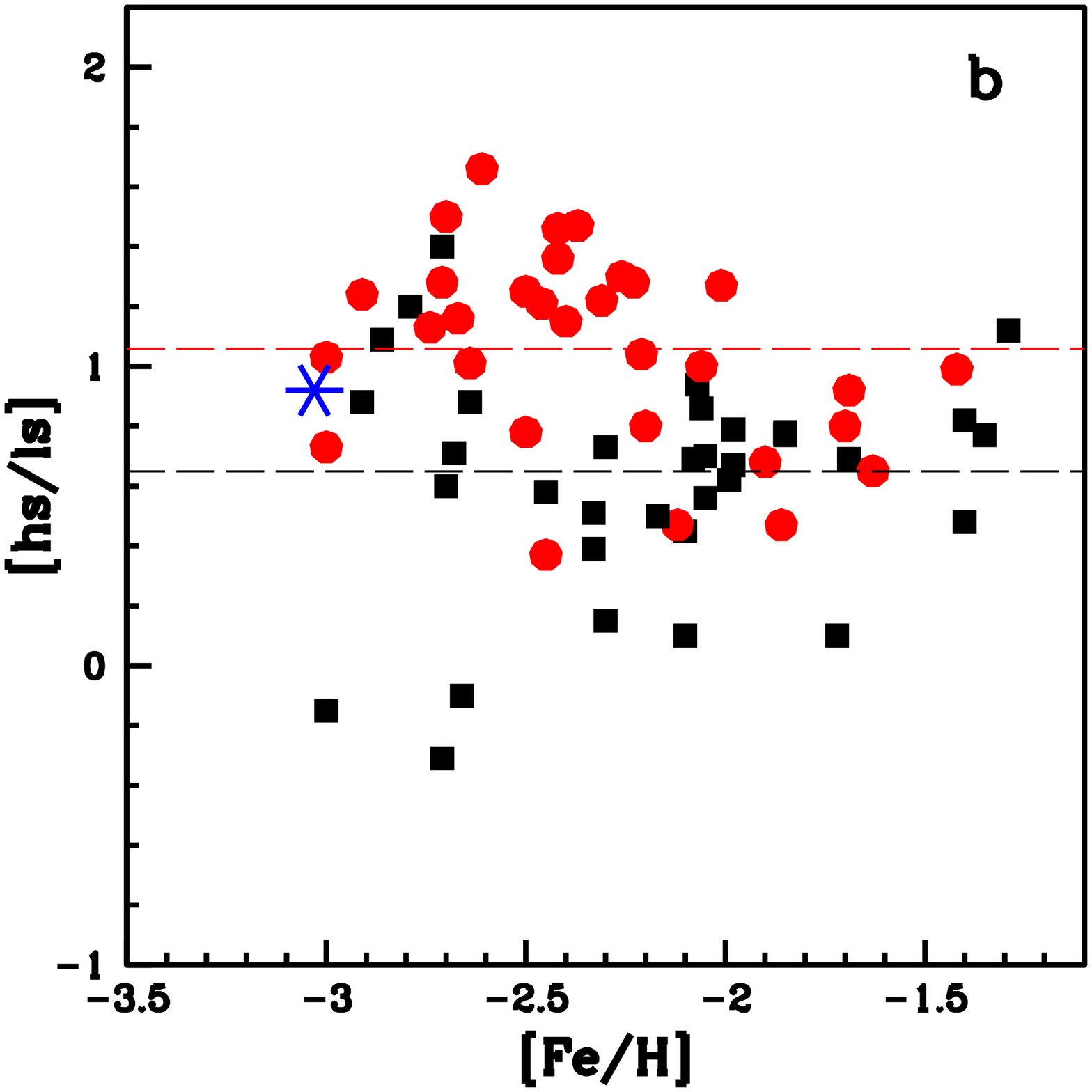}

        }
    \caption{Filled red circles and filled black squares respectively represent literature CEMP-r/s and CEMP-s stars compiled by \citet{Goswami_et_al_1_2021}, and the blue star represents the programme star. {\it{panel a:}} Grid formed by the dotted red lines bound by 0.0~$<$~[La/Eu]~$<$~0.6 and 0.0~$<$~[Ba/Eu]~$<$~1.0 indicates the region defined for CEMP-r/s stars by \citet{Goswami_et_al_1_2021}. The grid formed by the black dashed lines bound by 0.5~$<$~[La/Eu]~$<$~0.7 represents the region where [Eu/Fe]~$>$1.0 classifies the stars as CEMP-r/s and [Eu/Fe]~$<$1.0 classifies the stars as CEMP-s. {\it{panel b:}} Red dashed line at [hs/ls] = 1.06 and the black dashed line at [hs/ls] = 0.65 represent the peaks of [hs/ls] for CEMP-r/s and CEMP-s, respectively, as shown by \citet{Goswami_et_al_1_2021} in Figure~13(a).
     }%
   \label{fig:subfigures_classification}
       \end{center}
\end{figure*}

Our analysis  shows the object HE~1005--1439 to be an extremely metal-poor star
with [Fe/H]~=~--3.03 in accordance with \citet{aoki2007carbon}. 
Carbon and neutron-capture elements are found to be enhanced in the programme star. Following the classification criteria of CEMP stars  \citep{Goswami_et_al_1_2021}, the object is found to belong to the CEMP-s sub-group (Figure~\ref{fig:subfigures_classification}a). However, as shown in Figure~\ref{fig:subfigures_classification}b, the ratio of heavy-s process (hs) elements (Ba, La, Ce, and Nd) to the light-s process (ls) elements (Sr and Y), [hs/ls] ($\sim$ 0.92) is closer to the value at which  the CEMP-r/s stars peak (1.06) \citep{Goswami_et_al_1_2021}.

\par Using the {\it{s}}-process yields calculated using the FRUITY\footnote{\url{http://fruity.oa-teramo.inaf.it/}} model \citep{Straniero_et_al.2006, Cristallo_et_al.2009_I, Cristallo_et_al_2011, Cristallo_et_al_2015} at the same metallicity (z=0.00002) of the programme star, and considering different masses, we were not able to reproduce the observed abundance pattern of heavy elements. In Figure~\ref{fig:s_first}, we show a comparison of the observed elemental abundances  with  the AGB model yields (normalised to the La abundance of HE~1005--1439), calculated for M~=~1.3 M$_\odot$ and M~=~2.0 M$_\odot$. As can be seen in the top panel of the residual plot of Figure~\ref{fig:residue_fourth}, the {\it{s}}-process AGB models over-produce the light {\it{s}}-process elements Sr and Y, under-produce the elements Pr, Er, and Hf, and over-produce the third s-process peak element Pb.  

\par In Figure~\ref{fig:i_second}, we show a comparison of the {\it{i}}-process model yields of \citet{Hampel2016} (normalised to the La abundance of HE~1005--1439), calculated for n~=~10$^{12}$ cm$^{-3}$ and n~=~10$^{14}$ cm$^{-3}$, with the observed elemental abundance pattern of the programme star. The {\it{i}}-process models \citep{Hampel2016} alone at neutron-densities n~=~10$^{12}$ -- 10$^{15}$ cm$^{-3}$ cannot satisfactorily reproduce the observed abundance pattern of the programme star either. We can see from the middle panel of residual plot (Figure~\ref{fig:residue_fourth}) that the {\it{i}}-process model  with n~=~10$^{12}$ cm$^{-3}$ fits the light {\it{s}}-process elements Sr and Y satisfactorily but under-produces Ce and Pr. Again, {\it{i}}-process model with n~=~10$^{14}$ cm$^{-3}$ under produces the light {\it{s}}-process elements Sr and Y but over-produces Ba, Eu, and Er.
The diverse abundance pattern observed in HE~1005--1439, which could  not be explained either by  {\it{s}}-process AGB nucleosynthesis or by the {\it{i}}-process alone, 
prompted us to  explore alternate production mechanisms that might have influenced its surface chemical composition. 

\subsubsection{Parametric-model-based study}
\label{sec:parametric_model}

\begin{figure*}
     \begin{center}
\centering
        \subfigure[{\it{s}}-process model fits for M~=~1.3~M$_{\odot}$~\&~2.0~M$_{\odot}$]{%
            \label{fig:s_first}
            \includegraphics[height=8.0cm,width=8.5cm]{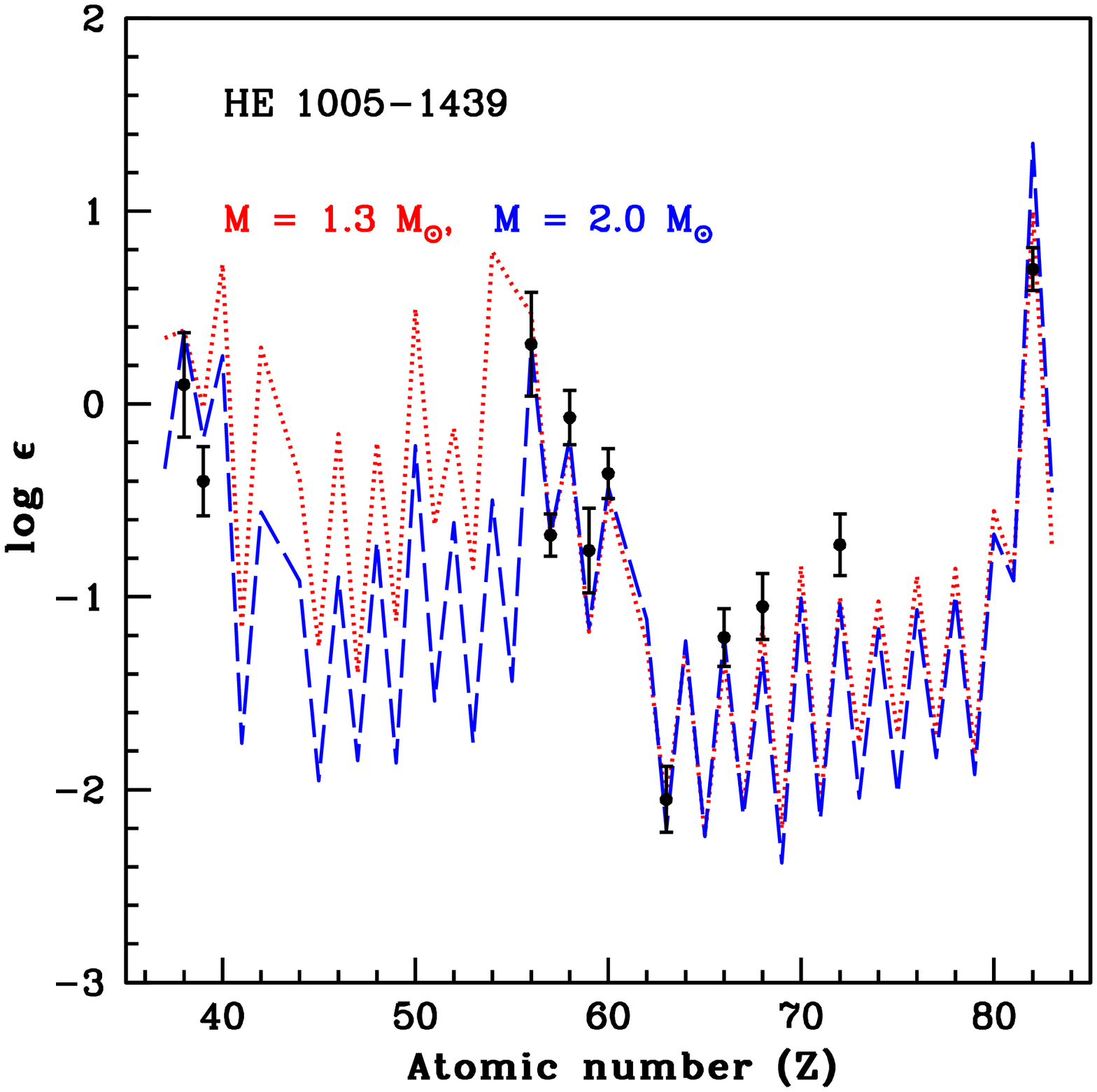}
        }%
        \subfigure[{\it{i}}-process model fits for n~=~10$^{12}$~cm$^{-3}$~\&~10$^{14}$~cm$^{-3}$ ]{%
            \label{fig:i_second}
            \includegraphics[height=8.0cm,width=8.5cm]{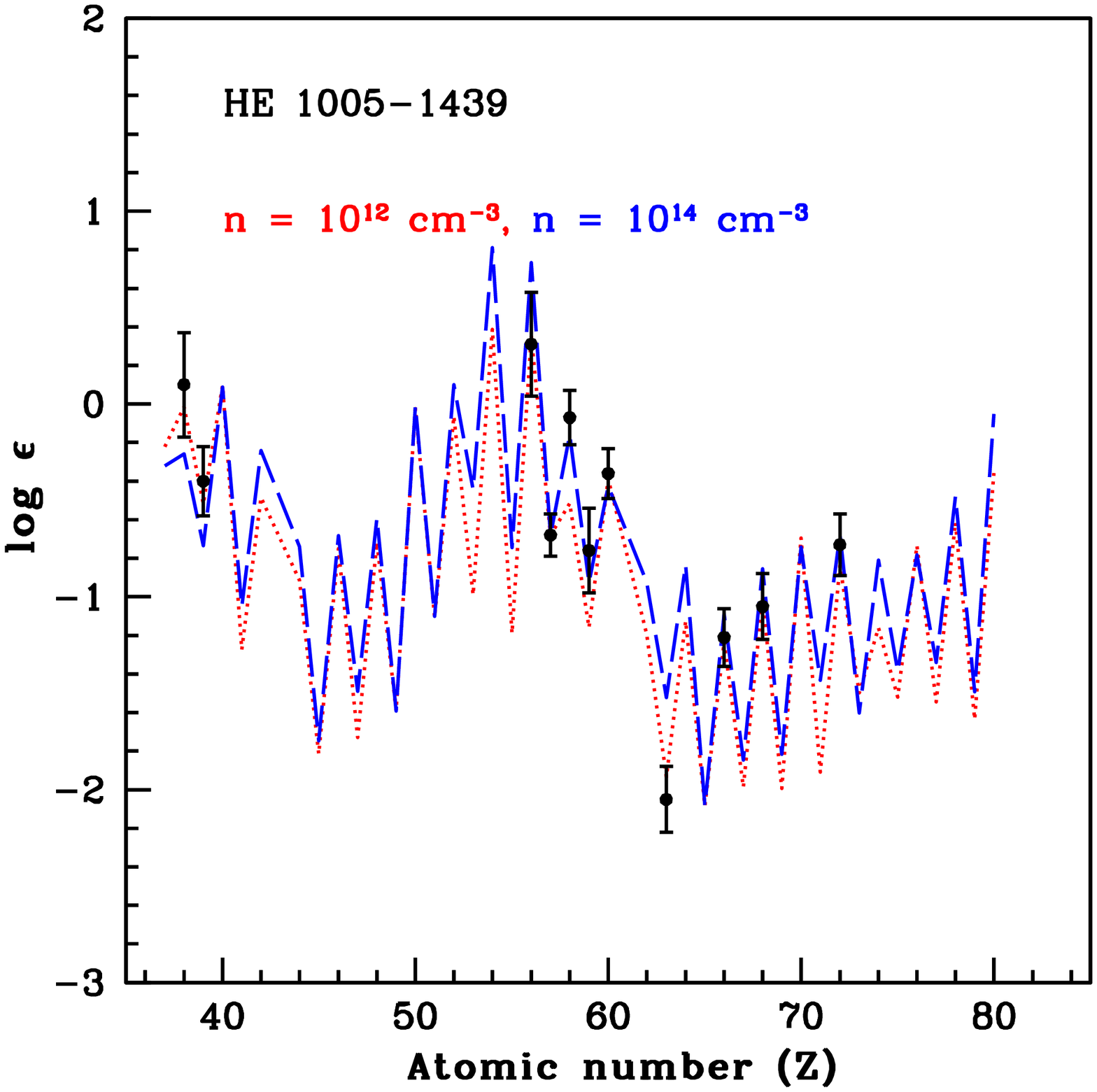}

        }\\ 

        \subfigure[Best-fit from parametric-model]{%
            \label{fig:best_third}
            \includegraphics[height=8.0cm,width=8.5cm]{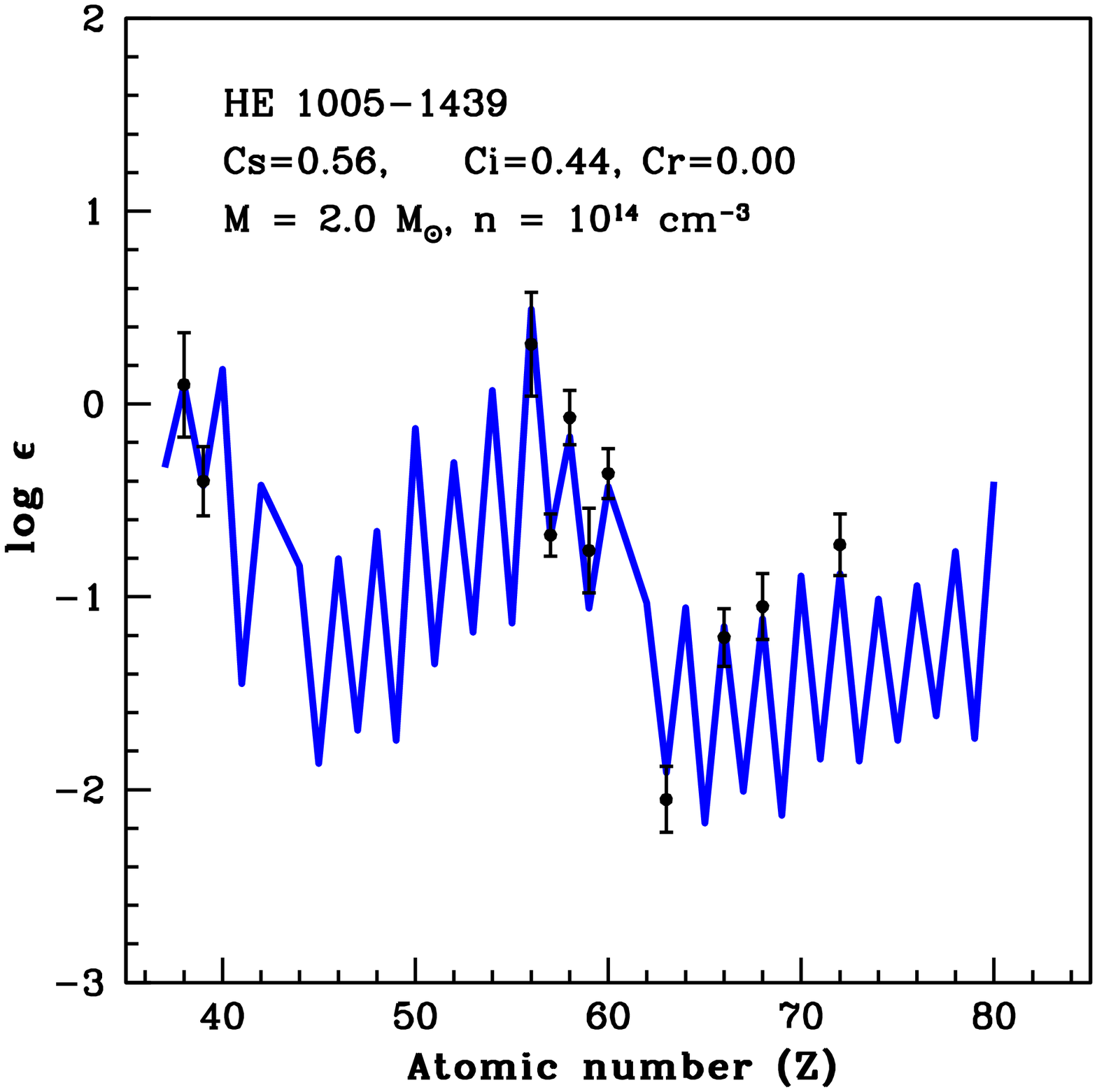}
        }%
        \subfigure[Residual plot]{%
            \label{fig:residue_fourth}
            \includegraphics[height=8.0cm,width=8.5cm]{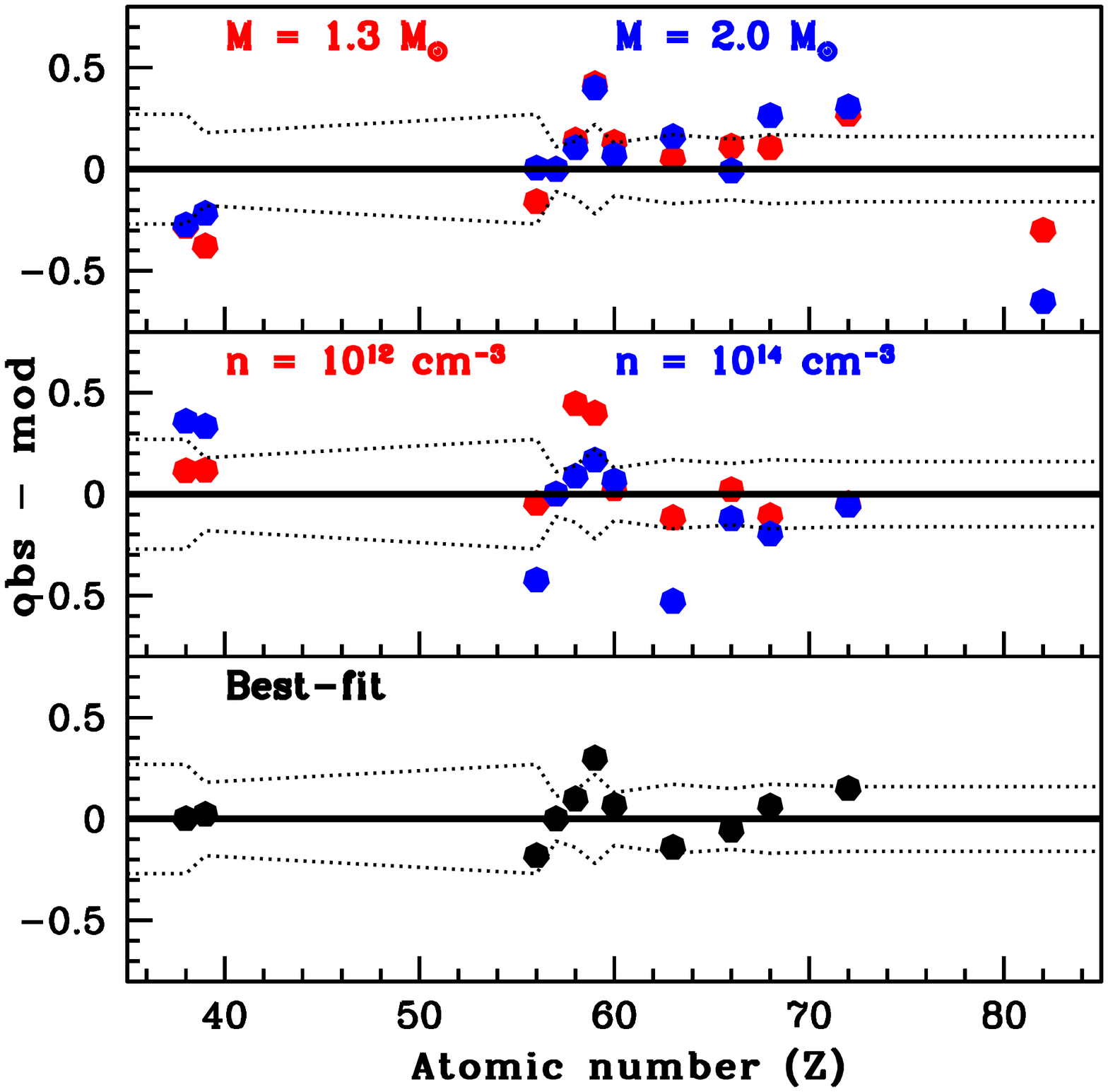}

        }\\ 

    \caption{ Examples of theoretical model fits with the observed abundances of the star. In panels (a),  (b),  and (c), the points with error bars indicate the observed abundances.}%
   \label{fig:fits}
       \end{center}

\end{figure*}

We performed a parametric-model-based study to delineate the contributions of {\it{s}}-, {\it{i}}- and {\it{r}}-processes to the observed abundances of heavy elements of the programme star. We used {\it{s}}-process model yields of the FRUITY model at different masses (M~=~1.3--2.0 M$_\odot$), the Solar System {\it{r}}-process residual pattern (stellar model) given in \citet{arlandini1999}, and {\it{i}}-process model yields of \citet{Hampel2016} at different neutron-densities (n~=~10$^{12}$--10$^{15}$ cm$^{-3}$). We excluded the element Pb from the parametric-model based study as the {\it{i}}-process model yields of Pb are not reported by \citet{Hampel2016}. We normalised the elemental abundances of the models to the La abundance of HE~1005--1439. The observed elemental abundances of HE~1005--1439 are then fitted with the parametric-model function log $\epsilon_{j}$~=~C$_{s}$ N$_{sj}$ + C$_{i}$N$_{ij}$+ C$_{r}$N$_{rj}$, where N$_{sj}$, N$_{ij,}$ and N$_{rj}$ indicate the normalised abundance from the {\it{s}}-process,  {\it{i}}-process, and  {\it{r}}-process, respectively. Here, C$_{s}$, C$_{i,}$ and C$_{r}$ indicate the component coefficients corresponding to contributions from the {\it{s}}-process, {\it{i}}-process, and {\it{r}}-process, respectively.  In order to find the best fit, we calculated $\chi^{2}$ for all the possible combinations of models of {\it{s}}-process (M~=~1.3--2.0 M$_\odot$) and {\it{i}}-process (n~=~10$^{12}$--10$^{15}$ cm$^{-3}$) along with solar {\it{r}}-process residues. The minimum  $\chi^{2}$ is achieved for a combination of an {\it{s}}-process model with M~=~2.0 M$_\odot$ and an {\it{i}}-process model with n~=~10$^{14}$ cm$^{-3}$ with no contribution from the {\it{r}}-process. The parametric-model function gives an excellent fit to the light s-process elements Sr and Y and a satisfactory fit to the heavier neutron-capture elements. The best parametric-model fit, where the contribution from both {\it{s}}- and {\it{i}}-process are similar (C$s$~=~0.56, C$i$~=~0.44), is shown in Figure~\ref{fig:best_third} and the bottom panel of the residual plot (Figure~\ref{fig:residue_fourth}).

\subsubsection{Origin of the programme star: Possible formation scenarios}
The parametric-model-based study clearly established that the surface chemical composition of HE 1005$-$1439 is influenced by almost equal contributions of {\it{s}}-process AGB nucleosynthesis and  {\it{i}}-process nucleosynthesis. We attempted to capture a formation scenario for this object involving a binary picture. Our assumption on binarity is profoundly based  on the radial velocity variations observed on a few epochs. The low value of $^{12}$C/$^{13}$C ($\sim$ 5.0) measured for HE~1005--1439 also points towards the extrinsic nature of carbon, and hence the heavy elements in the star. In the intrinsic carbon stars (as they are in AGB phase), third-dredge-up (TDU) episodes bring $^{12}$C and {\it{s}}-process material to the surface, and the $^{12}$C/$^{13}$C ratio can increase up to $>$ 100 depending on the initial mass of the object \citep{Karakas_&_Lattanzio_2014}. However, in extrinsic carbon giants, such as HE~1005--1439, the $^{12}$C/$^{13}$C ratio decreases due to the first dredge-up (FDU), which brings $^{13}$C produced in the internal CNO cycle to the stellar atmosphere \citep{smith_et_al_1993}. Attributing the s-process contribution to the AGB mass-transfer scenario in the binary system, we carefully examined if any of the proposed formation scenarios of {\it{i}}-process nucleosynthesis available in literature could explain the {\it{i}}-process contribution in the observed 
abundance pattern of HE~1005$-$1439.

\par We investigated 
if the very late thermal pulse (VLTP) scenario proposed to explain the peculiar abundance 
pattern of the Sakurai object (V4334 Sagittarii) by \citet{Herwig_et_al_2011} could explain the abundance peculiarities of HE~1005$-$1439. However, a characteristic 
property of VLTP in pre-white dwarf of producing light {\it{s}}-process elements
2 dex more than that of heavy {\it{s}}-process elements is not observed in 
HE~1005$-$1439. 

\par  Low-metallicity or zero-metallicity massive (20--30 M$_\odot$) stars \citep{Bannerjee_et_al_2018} and super-AGB stars (9--11 M$_\odot$) \citep{Jones_et_al_2016} can pollute the ISM with {\it{i}}-process yields. However, an AGB mass-transfer scenario in a binary system formed from the {\it{i}}-process-enriched ISM also cannot explain the peculiar abundance pattern of the programme star due to its inadequacy to describe the low observed Pb abundance in HE~1005--1439. As shown in Figure~\ref{fig:s_first} and the top panel of Figure~\ref{fig:residue_fourth}, the observed Pb abundance of HE~1005--1439 is about 0.30--0.60 dex lower than the prediction of the FRUITY model \citep{Cristallo_et_al.2009_I, Cristallo_et_al_2016} at [Fe/H]~$\sim$ --3.0. Pb is the main product of the {\it{s}}-process in AGB stars at low metallicities \citep{Cristallo_et_al.2009_I}. The low Pb abundance of HE~1005--1439 is difficult to explain by the scenario of pre-enrichment of ISM with {\it{i}}-process material or a scenario involving any external source of {\it{i}}-process material such as rapidly accreting white dwarfs \citep{Denissenkov_et_al_2019}. This is because in these cases Pb would be more than (if the {\it{i}}-process neutron exposure ($\tau$) of the progenitor is sufficient to produce enough Pb) or at a similar level to the {\it{s}}-process prediction, but it would not be less, as Pb is not destroyed in $\beta$-decay.

\par We propose that  the surface chemical composition of HE~1005--1439 may be attributed to mass-transfer from a now extinct AGB 
companion with  both {\it{s}}- and {\it{i}}-process nucleosynthesis occurring under suitable conditions  during its evolution at  different thermal pulses. Such a formation scenario may not be unlikely as many studies  on the evolution of AGB stars 
have shown that neutron densities required for {\it{s}}-process and {\it{i}}-process can be achieved in the intershell region with the help of the partial mixing of protons in the radiative conditions \citep{Herwig_2000, Denissenkov_&_Tout_2003, Herwig_et_al_2003, Cristallo_et_al.2009_I, Cristallo_et_al_2011, Piersanti_et_al_2013, Karakas_&_Lattanzio_2014} and efficient PIEs in the convective conditions \citep{ Hollowell_et_al_1990, Fujimoto_et_al_2000, Iwamoto_et_al_2004, Campbell_&_Lattanzio_2008, Lau_et_al_2009, Cristallo_et_al_2009, Cristallo_et_al_2016, Choplin_et_al_2021}, respectively.

In a certain study \citep{Cristallo_et_al_2009}  on  the AGB evolution of a 1.5 M$_{\odot}$ model with [Fe/H]~=~--2.45 without $\alpha$ enhancement, it was found that  a strong PIE is followed by a  deep TDU. Due to the PIE, H-burning occurring in high-temperature convective conditions creates a huge amount of $^{13}C$, which leads to an efficient neutron production with neutron densities of the order of 10$^{15}$ cm$^{-3}$. The study also noted that the convective He-shell splits into two sub-shells when the energy released by proton capture reactions slightly exceeds the energy production at the base of the convective shell. In the lower sub-shell, the $^{13}C$~($\alpha$,~n)~$^{16}O$ reaction produces {\it{i}}-process neutron density, and the nucleosynthesis path goes away from the valley of $\beta$ stability producing a very high local [hs/ls] ratio. The upper sub-shell, where the CNO cycle is the main energy source, later gets engulfed by the envelope. After that, a standard TP-AGB phase follows with repetitive TPs and TDUs, and {\it{s}}-process nucleosynthesis occurs.  
 The number of expected stars experiencing PIEs is significantly reduced with the introduction of $\alpha$-element enhancements \citep{Cristallo_et_al_2016}.

\citet{Cristallo_et_al_2009} noted that the minimum mass of the models to experience TDU is significantly lowered by PIEs. Their model, after the PIE, encounters 25 additional TDU episodes, each of which is followed by radiative burning of the $^{13}C$ pocket. Although the final abundance pattern is a combination of {\it{i}}- and {\it{s}}-processes,   excessive {\it{s}}-process nucleosynthesis after the PIE would remove the trace of the {\it{i}}-process and thus reduce the [hs/ls] ratio.
 
In a recent study on the evolution of a 1 M$_{\odot}$ object at low metallicity  ([Fe/H]~=~--2.5), \citet{Choplin_et_al_2021} noticed three convective instabilities occurring at the beginning of the TP-AGB phase. 
The main PIE occurs during the third instability, which produces high neutron densities of about 4.3~$\times$~10$^{14}$ cm$^{-3}$ and rich {\it{i}}-process nucleosynthesis occurs.
We note that although they are of similar  metallicities,
 while in the model of \citet{Choplin_et_al_2021} no further TPs were possible after the main PIE, in that of \citet{Cristallo_et_al_2009} 25 TPs after the PIE masked the contribution of the {\it{i}}-process and resulted in an {\it{s}}-process surface abundance pattern. 

\par We propose that a model that undergoes PIEs during the beginning of the  TP-AGB phase (producing an {\it{i}}-process abundance pattern) followed by limited TPs (producing an {\it{s}}-process abundance pattern) might explain the  abundance peculiarity of HE~1005--1439. This scenario is also likely to explain the low-Pb abundance in the programme star. The abundance of Pb depends on the time-integrated neutron-exposure ($\tau$). From Figure~1 of \citet{Hampel_2019}, it is clear that as neutron density increases, production of Pb starts at higher $\tau$. Therefore, a low Pb abundance may indicate that due to the operation of both {\it{i}}- and {\it{s}}-processes in succession, neither of the processes received sufficient exposure time to produce enough Pb expected at [Fe/H] $\sim$ --3.0.

\section{Conclusions}
\label{sec:conclusion}
A detailed, high-resolution spectroscopic study of the EMP star HE~1005--1439 revealed a peculiar abundance pattern  different  from those typically  exhibited by CEMP stars. While the  CEMP stars' classification criteria place the object in the CEMP-s group, the value of [hs/ls] ${\sim}$ 0.9 is closer to the value (1.06) at which CEMP-r/s stars peak. The  abundance pattern could not be reproduced either by the {\it{s}}-process or the  {\it{i}}-process  model predictions alone.  However, a  parametric-model based analysis  clearly indicated that similar contributions from both the {\it{s}}- and {\it{i}}-process might have resulted in  the observed abundance pattern of HE~1005--1439.  We  propose that the origin of the observed  peculiar abundance pattern may be attributed to mass transfer from  a now extinct AGB companion where both {\it{i}}- and {\it{s}}-process nucleosynthesis took place during various stages of the AGB evolution with PIEs triggering {\it{i}}-process  followed by {\it{s}}-process AGB nucleosynthesis with a few TDU episodes.  

\par Several uncertainties such as initial mass, metallicity, treatment of opacities, nuclear rates, and mixing mechanisms affect the theoretical understanding of AGB stars. The observational constraints derived from  the programme star and the proposed scenario might provide a link for a better understanding of  the interplay between PIEs and partial mixing of protons in the intershell region and also the conditions resulting in a pure {\it{s}}- or {\it{i}}-process surface abundance pattern in low-mass, low-metallicity AGB stars. We believe that the proposed scenario will be helpful in explaining the overlap of [hs/ls] ratio in CEMP-s and CEMP-r/s stars (see Fig~13 of \citet{Goswami_et_al_1_2021}) and  the smooth transition of elemental abundances from the CEMP-s to CEMP-r/s regime \citep{Goswami_et_al_1_2021}.\\

\begin{acknowledgements}
Funding from DST SERB project EMR/2016/005283 is gratefully acknowledged. We thank Melanie Hampel for the {\it{i}}-process yields. We thank the referee Elisabetta Caffau for many constructive suggestions and useful comments. This work made use of the SIMBAD astronomical database, operated at CDS, Strasbourg, France, the NASA ADS, USA and data from the European Space Agency (ESA) mission Gaia (\url{https://www.cosmos.esa.int/gaia}), processed by the Gaia Data Processing and Analysis Consortium (DPAC, \url{https://www.cosmos.esa.int/web/gaia/dpac/consortium}).
\end{acknowledgements}

\bibliographystyle{aa}
\bibliography{sample}

\onecolumn
\begin{appendix} 
\section{Line list}
{\footnotesize
\begin{table}[H]
\caption{\bf{Equivalent widths (in m{\rm \AA}) of lines used for the calculation of elemental abundances.}}
\label{tab:Elem_linelist1} 
\scalebox{0.87}{
\begin{tabular}{ccccc}
\hline
Wavelength   &Element    &E$_{low}$ &   log gf  &  HE~1005--1439      \\
(\r{A})      &           & (eV)     &           &                 \\
\hline 
4132.06      &  Fe I     &  1.61    & $-$0.650  &  77.1 (4.40)    \\
4143.87      &           &  1.56    & $-$0.450  &  77.7 (4.16)    \\
4147.67      &           &  1.48    & $-$2.104  &  28.5 (4.56)    \\
4187.04      &           &  2.45    & $-$0.548  &  49.0 (4.51)    \\
4202.03      &           &  1.48    & $-$0.708  &  91.1 (4.71)    \\
4216.18      &           &  0.00    & $-$3.356  &  35.9 (4.28)    \\
4227.43      &           &  3.33    &    0.230  &  33.5 (4.34)    \\
4233.60      &           &  2.48    & $-$0.604  &  38.1 (4.35)    \\
4250.12      &           &  2.47    & $-$0.405  &  62.0 (4.67)    \\
4250.79      &           &  1.56    & $-$0.710  &  81.9 (4.49)    \\
4260.47      &           &  2.40    & $-$0.020  &  64.2 (4.26)    \\
4375.93      &           &  0.00    & $-$3.031  &  70.7 (4.66)    \\
4383.55      &           &  1.48    &    0.200  & 127.1 (4.70)    \\
4476.02      &           &  2.85    & $-$0.570  &  30.1 (4.52)    \\
4528.61      &           &  2.18    & $-$0.822  &  62.0 (4.71)    \\ 
4602.94      &           &  1.48    & $-$1.950  &  37.3 (4.53)    \\
4872.14      &           &  2.88    & $-$0.600  &  25.7 (4.45)    \\
4890.76      &           &  2.88    & $-$0.430  &  39.2 (4.56)    \\
4891.49      &           &  2.85    & $-$0.140  &  51.8 (4.49)    \\
4918.99      &           &  2.87    & $-$0.370  &  35.6 (4.42)    \\
4920.50      &           &  2.83    &    0.060  &  55.7 (4.36)    \\
4994.13      &           &  0.92    & $-$3.080  &  19.7 (4.58)    \\
5006.12      &           &  2.83    & $-$0.615  &  23.4 (4.34)    \\
5171.60      &           &  1.49    & $-$1.793  &  46.9 (4.50)    \\
5192.34      &           &  3.00    & $-$0.421  &  27.1 (4.41)    \\
5194.94      &           &  1.56    & $-$2.090  &  29.4 (4.53)    \\
5216.27      &           &  1.61    & $-$2.150  &  25.6 (4.56)    \\
5226.86      &           &  3.04    & $-$0.555  &  19.7 (4.39)    \\
5227.19      &           &  1.56    & $-$1.228  &  59.1 (4.24)    \\
5232.94      &           &  2.94    & $-$0.190  &  34.4 (4.27)    \\
5266.56      &           &  3.00    & $-$0.490  &  29.8 (4.54)    \\
5269.54      &           &  0.86    & $-$1.321  &  89.4 (4.16)    \\
5270.36      &           &  1.61    & $-$1.510  &  65.8 (4.71)    \\
5324.18      &           &  3.21    & $-$0.240  &  31.0 (4.54)    \\
5328.04      &           &  0.91    & $-$1.466  &  81.8 (4.18)    \\
5328.53      &           &  1.56    & $-$1.850  &  35.7 (4.41)    \\
5446.92      &           &  0.99    & $-$1.930  &  63.2 (4.34)    \\
5455.61      &           &  1.01    & $-$2.091  &  60.1 (4.46)    \\
5497.52      &           &  1.01    & $-$2.849  &  26.7 (4.58)    \\
5506.78      &           &  0.99    & $-$2.797  &  25.0 (4.46)    \\
5586.76      &           &  3.37    & $-$0.210  &  25.5 (4.54)    \\
5615.64      &           &  3.33    & $-$0.140  &  29.0 (4.50)    \\
6136.61      &           &  2.45    & $-$1.400  &  15.6 (4.41)    \\
6230.72      &           &  2.56    & $-$1.281  &  16.7 (4.44)    \\
6494.98      &           &  2.40    & $-$1.273  &  26.0 (4.48)    \\
4491.40      &  Fe II    &  2.86    & $-$2.700  &   7.6 (4.53)    \\
4923.93      &           &  2.89    & $-$1.320  &  51.3 (4.46)    \\
5018.44      &           &  2.89    & $-$1.220  &  56.5 (4.48)    \\
5276.00      &           &  3.20    & $-$1.940  &  16.2 (4.50)    \\
5316.61      &           &  3.15    & $-$1.850  &  20.9 (4.51)    \\
5889.95      &  Na I     &  0.00    &    0.117  & 161.2 (4.56)    \\
5895.92      &           &  0.00    & $-$0.184  & 144.2 (4.60)    \\
4571.10      &  Mg I     &  0.00    & $-$5.691  &  16.0 (5.03)    \\
5172.68      &           &  2.71    & $-$0.402  & 125.9 (5.12)    \\
5183.60      &           &  2.72    & $-$0.180  & 138.4 (5.14)    \\
5528.40      &           &  4.35    & $-$0.620  &  33.4 (5.08)    \\
4226.73      &  Ca I     &  0.00    &    0.243  & 144.6 (4.00)    \\
4318.65      &           &  1.90    & $-$0.208  &  22.1 (3.55)    \\
4435.68      &           &  1.89    & $-$0.500  &  26.9 (3.95)    \\
4585.86      &           &  2.53    & $-$0.186  &  10.4 (3.77)    \\
5265.56      &           &  2.52    & $-$0.260  &  13.4 (3.94)    \\
5588.75      &           &  2.53    &    0.210  &  25.8 (3.84)    \\
5594.46      &           &  2.52    & $-$0.050  &  20.0 (3.93)    \\
6122.22      &           &  1.89    & $-$0.409  &  23.3 (3.68)    \\
6162.17      &           &  1.90    &    0.100  &  36.4 (3.50)    \\
6439.07      &           &  2.53    &    0.470  &  28.9 (3.63)    \\
4246.82      &  Sc II    &  0.32    &    0.320  & 100.6 (0.89)    \\
4415.56      &           &  0.60    & $-$0.640  &  54.2 (0.75)    \\
4981.73      &  Ti I     &  0.85    &    0.504  &  24.1 (2.31)    \\
4991.06      &           &  0.84    &    0.380  &  12.6 (2.06)    \\
5007.21      &           &  0.82    &    0.112  &  17.5 (2.48)    \\
5210.39      &           &  0.05    & $-$0.884  &   8.5 (2.24)    \\

\hline
\end{tabular}}
\end{table}}

{\footnotesize
\begin{table}
{\bf{\it{-continued\\}}}
\scalebox{0.87}{
\begin{tabular}{ccccc}
\hline
Wavelength   &Element    &E$_{low}$ &   log gf  &  HE~1005--1439      \\
(\r{A})      &           & (eV)     &           &                 \\
\hline 
4290.22      &  Ti II    &  1.16    & $-$1.120  &  63.8 (2.54)    \\
4395.03      &           &  1.08    & $-$0.660  &  75.7 (2.33)    \\
4450.48      &           &  1.08    & $-$1.450  &  28.5 (1.90)    \\
4468.51      &           &  1.13    & $-$0.600  &  72.3 (2.19)    \\
4563.76      &           &  1.22    & $-$0.960  &  63.3 (2.38)    \\
4805.09      &           &  2.06    & $-$1.100  &  16.6 (2.28)    \\
5188.68      &           &  1.58    & $-$1.210  &  31.0 (2.21)    \\
5226.54      &           &  1.57    & $-$1.300  &  20.2 (2.02)    \\
4254.34      &  Cr I     &  0.00    & $-$0.114  &  73.4 (2.60)    \\
4274.80      &           &  0.00    & $-$0.231  &  76.3 (2.80)    \\
5204.51      &           &  0.94    & $-$0.208  &  39.4 (2.66)    \\
5206.04      &           &  0.94    &    0.019  &  39.9 (2.44)    \\
5208.43      &           &  0.94    &    0.158  &  46.7 (2.46)    \\
4030.75      &  Mn I     &  0.00    & $-$0.470  &  81.7 (2.66)    \\
4033.06      &           &  0.00    & $-$0.618  &  72.1 (2.47)    \\
4092.38      &  Co I     &  0.92    & $-$0.940  &  14.9 (2.23)    \\
4118.77      &           &  1.05    & $-$0.490  &  25.9 (2.24)    \\
4121.31      &           &  0.92    & $-$0.320  &  33.3 (2.12)    \\
5476.90      &  Ni I     &  1.83    & $-$0.890  &  33.0 (3.40)    \\
4077.71      &  Sr II    &  0.00    &    0.167  &  96.7 ($-$0.05)    \\
4215.52      &           &  0.00    & $-$0.145  &  98.9 (0.26)    \\
4883.68      &  Y II     &  1.08    &    0.070  &  15.6 ($-$0.55)    \\
4900.12      &           &  1.03    & $-$0.090  &  12.5 ($-$0.57)    \\
5087.42      &           &  1.08    & $-$0.170  &  23.0 ($-$0.10)    \\
5853.67      &  Ba II    &  0.60    & $-$1.000  &  52.5 (0.37)    \\
6141.71      &           &  0.70    & $-$0.076  &  89.2 (0.65)    \\
6496.90      &           &  0.60    & $-$0.377  &  93.5 (0.90)    \\
4086.71      &  La II    &  0.00    & $-$0.150  &  31.6 ($-$0.69)    \\
4123.22      &           &  0.32    &    0.120  &  29.4 ($-$0.66)    \\
4921.78      &           &  0.24    & $-$0.680  &  15.2 ($-$0.48)    \\
4137.65      &  Ce II    &  0.52    &    0.246  &  14.3 ($-$0.26)    \\
4486.91      &           &  0.30    & $-$0.474  &  11.5 (0.05)    \\
4562.36      &           &  0.48    &    0.081  &  20.5 (0.02)    \\
4179.39      &  Pr II    &  0.20    &    0.310  &  15.7 ($-$0.76)    \\
4061.08      &  Nd II    &  0.47    &    0.550  &  31.2 ($-$0.30)    \\
4109.07      &           &  0.06    &    0.280  &  31.6 ($-$0.49)    \\
4156.08      &           &  0.18    &    0.200  &  25.0 ($-$0.46)    \\
4451.56      &           &  0.38    & $-$0.040  &  20.5 ($-$0.17)    \\
3944.68      &  Dy II    &  0.00    &    0.030  &  10.9 ($-$1.27)    \\
4000.45      &           &  0.10    &    0.009  &  14.3 ($-$1.00)    \\
4077.97      &           &  0.10    & $-$0.058  &   6.4 ($-$1.36)    \\
3896.23      &  Er II    &  0.05    & $-$0.243  &  13.9 ($-$0.91)    \\
3906.31      &           &  0.00    & $-$0.052  &  13.0 ($-$1.20)    \\
3793.38      &  Hf II    &  0.38    & $-$0.950  &   6.7 ($-$0.72)    \\
3918.09      &           &  0.45    & $-$1.010  &   5.0 ($-$0.74)    \\
\hline
\end{tabular}}

The numbers in parentheses in column 5 give the derived abundances from the respective lines.\\. 
\end{table}}

\end{appendix}

\label{lastpage}
\end{document}